 \documentclass[final,5p,times,twocolumn,numbers,sort&compress]{elsarticle}

\usepackage{amssymb}
\usepackage{lipsum}
\usepackage{amsmath}
\usepackage{bbm} 
\usepackage{bm}
\usepackage{xcolor}
\usepackage{soul}

\newcommand{\Mcal}{\mathcal{M}}

\newcommand{\Ccal}{\mathcal{C}}

\newcommand{\vect}[1]{\boldsymbol{#1}}
\newcommand{\kt}{\vect{k}}

\newcommand{\Pt}{\vect{P}}

\newcommand{\qt}{\vect{q}}
\newcommand{\lt}{\vect{\ell}}

\newcommand{\bt}{\vect{b}}

\newcommand{\Kt}{\vect{K}}

\newcommand{\qtone}{\boldsymbol{q_{1}}}
\newcommand{\qttwo}{\boldsymbol{q_{2}}}

\newcommand{\xt}{\vect{x}}
\newcommand{\yt}{\vect{y}}
\newcommand{\zt}{\vect{z}}

\newcommand{\rt}{\vect{r}}

\newcommand{\ellt}{\boldsymbol{\ell}}
\newcommand{\Deltat}{\boldsymbol{\Delta}}

\newcommand{\et}{\vect{\epsilon}}

\newcommand{\der}{\mathrm{d}}

\newcommand{\Tr}{\mathrm{Tr}}

\newcommand{\eqn}[1]{Eq.\,\eqref{#1}}
\newcommand{\beq}{\begin{equation}}
\newcommand{\eeq}{\end{equation}}
\newcommand{\bal}{\begin{align}}
\newcommand{\eal}{\end{align}}

\newcommand{\br}{\bm{r}}

\newcommand{\KT}{K_\perp}
\newcommand{\PT}{P_\perp}

\journal{Physics Letters B}

\begin{document}

\begin{frontmatter}



\title{Unveiling the sea: universality of the transverse momentum dependent quark distributions at small $x$}


\author[a]{Paul Caucal}
 \author[b]{Marcos Guerrero Morales}
 \author[c]{Edmond Iancu}
 \author[b,d,e,f]{Farid Salazar}
 \author[c,g]{Feng Yuan}
 \affiliation[a]{SUBATECH UMR 6457 (IMT Atlantique, Universite de Nantes, IN2P3-CNRS), 4 rue Alfred Kastler, 44307 Nantes, France}
\affiliation[b]{Department of Physics, Temple University, Philadelphia, PA 19122 - 1801, USA}
 \affiliation[c]{Universite Paris-Saclay, CNRS, CEA, Institut de physique theorique, F-91191, Gif-sur-Yvette, France}
 \affiliation[d]{RIKEN-BNL Research Center, Brookhaven National Laboratory, Upton, New York 11973, USA}
 \affiliation[e]{Physics Department, Brookhaven National Laboratory, Upton, New York 11973, USA}
 \affiliation[f]{Institute for Nuclear Theory, University of Washington, Seattle WA 98195-1550, USA}
\affiliation[g]{Nuclear Science Division, Lawrence Berkeley National Laboratory, Berkeley, CA 94720, USA}

\begin{abstract}
Within the Colour Glass Condensate effective theory, we demonstrate that  back-to-back dijet correlations in dilute-dense collisions involving a small-$x$ quark from the nuclear target  can be factorised in terms of universal transverse momentum dependent distributions (TMDs) for the sea quarks. Two building blocks are needed to construct all these  TMDs  at the operator level: the sea quark TMD operator which appears in semi-inclusive Deep-Inelastic Scattering (SIDIS) or in the Drell-Yan process and the elastic $S$-matrix for a quark-antiquark dipole. Compared to SIDIS, the saturation effects are stronger for dijet production in forward proton-nucleus collisions, due to additional scattering in the initial and final state, effectively resulting in a larger value for the nuclear saturation momentum.
\end{abstract}




\end{frontmatter}

\section{Introduction}

One of the key insights from the high-energy electron-proton collisions at HERA is the rapid rise of the gluon distribution in protons as the longitudinal momentum fraction $x$ of the gluon with respect to the proton decreases~\cite{ZEUS:1993ppj}. This steep growth is moderated by non-linear gluon saturation~\cite{Gribov:1984tu,Mueller:1985wy,McLerran:1993ni}, an emergent phenomenon described by the Colour Glass Condensate (CGC) effective theory~\cite{Iancu:2002xk,Iancu:2003xm,Gelis:2010nm,Kovchegov:2012mbw,Morreale:2021pnn}. The experimental characterisation of gluon saturation is a central objective of the future Electron-Ion Collider (EIC)~\cite{Accardi:2012qut,AbdulKhalek:2021gbh,Achenbach:2023pba}, which will extend the groundbreaking work of HERA~\cite{Stasto:2000er,Marquet:2006jb}. On the other hand, HERA also showed~\cite{ZEUS:1998agx,H1:1997uzt} that the proton wave-function at small $x$ contains a substantial component of ``sea" quarks, which arise primarily from quark-antiquark pair production from small-$x$ gluons. In this Letter, we demonstrate that the CGC consistently accounts for the sea quark contribution~\cite{McLerran:1998nk,Mueller:1999wm,Venugopalan:1999wu} inside nuclei and predicts the shape of the quark transverse momentum dependent (TMD)  distributions~\cite{Collins:2011zzd,Boussarie:2023izj} in the saturation regime.

Our analysis builds upon pioneering work which unveiled sea quark TMD factorisation for selected processes in  electron-nucleus  ($eA$) deep inelastic scattering (DIS) at small $x$: the semi-inclusive production (SIDIS) of quark jets (or hadrons)~\cite{Marquet:2009ca,Xiao:2017yya,Caucal:2024vbv}, lepton-jet correlations~\cite{Tong:2022zwp}, Drell-Yan \cite{Kang:2012vm,Gelis:2006hy}, and the diffractive production of quark jets \cite{Hatta:2022lzj} and of quark-gluon dijets \cite{Hauksson:2024bvv}. Despite such interesting, but rare, examples, a systematic understanding of the emergence of quark TMD factorisation at small $x$ --- similar to that for the case of gluons~\cite{Dominguez:2010xd,Dominguez:2011wm,Dominguez:2011br,Metz:2011wb,Dumitru:2015gaa,Kotko:2015ura,Marquet:2016cgx,Marquet:2017xwy,Stasto:2018rci,Albacete:2018ruq,Dumitru:2018kuw,Altinoluk:2018byz,Mantysaari:2019hkq,Altinoluk:2019wyu,Boussarie:2020fpb,Altinoluk:2020qet,Boussarie:2021ybe,Iancu:2022lcw,Benic:2022ixp,Deganutti:2023qct,Ganguli:2023joy,vanHameren:2023oiq,Cheung:2024qvw,Kotko:2017oxg,Iancu:2021rup,Caucal:2024bae} --- was still missing. It is our purpose in this Letter to fill this gap. Via the study of representative processes describing 
inclusive two-particle production (dijets, dihadrons, or photon-jets, to be collectively referred to as ``dijets'' in what follows) in $eA$ DIS and proton-nucleus ($pA$) collisions, we will demonstrate the ubiquitousness and the universality of the sea quark TMDs, as computed from the CGC.

The dijet processes of interest for us here involve the scattering between a parton from the dilute projectile and a sea quark with $x\ll 1$ from  the nuclear  target, hence they are suppressed by a power of the QCD coupling $\alpha_s$ compared to gluon-mediated processes leading to similar final states.
Indeed, sea quarks are produced in pairs, via the gluon splitting $g\to q\bar q$, hence their distribution is proportional to that of the small-$x$ gluons, with a proportionality constant of order $\alpha_s$. This argument also shows that processes involving the sea quark distribution show exactly the same behaviour at high energy as their gluonic counterparts. So, their contributions must be taken into account for the phenomenology of back-to-back two-particle correlations at high energies, as probed via  forward particle production in $pA$ collisions  at the Large Hadron Collider, or in  $eA$ collisions at the EIC. Since the sea quark TMD is sensitive to gluon saturation~\cite{McLerran:1998nk,Mueller:1999wm,Venugopalan:1999wu} (see also below), our results open a new window for saturation studies at colliders through the quark channel.

\begin{figure}
    \centering
    \includegraphics[width=0.38\linewidth]{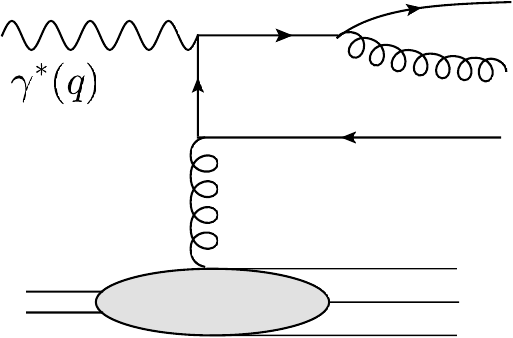} \qquad \includegraphics[width=0.42\linewidth]{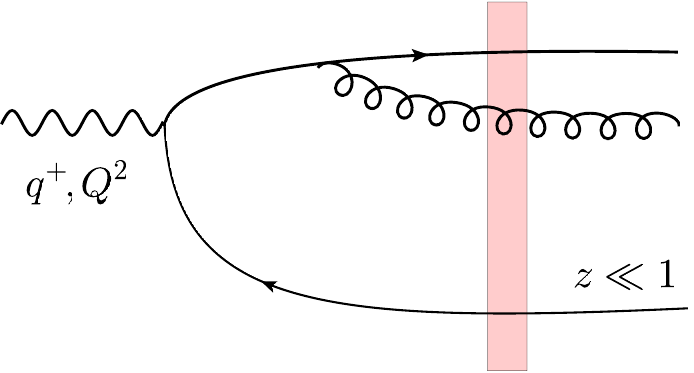} 
    \caption{Example of graphs contributing to quark-gluon production in DIS at small $x$ and to leading order in $\alpha_s$. Left: target picture  (the photon is absorbed by a sea quark, which then decays into a quark-gluon pair). Right: CGC picture
     (a quark-antiquark-gluon fluctuation of the photon scatters off the shockwave target field; the soft ($z\ll 1$) antiquark is not measured). }
    \label{fig:seaQ}
\end{figure}

To be more specific, consider quark-gluon dijet production in DIS, as illustrated (within the standard parton picture for the target) by the left Feynman graph in Fig~\ref{fig:seaQ}. The dijets are  assumed to be hard and nearly back-to-back in the transverse plane: their relative transverse momentum $P_\perp$  is much larger than both their momentum imbalance $K_\perp$ and the target saturation momentum $Q_s(x)$. Despite being hard, the dijets can probe low-$x$ partons in the target wavefunction provided the center-of-mass (COM) energy $\sqrt{\hat s}$ for the photon-nucleon collision is sufficiently high; one parametrically has $x\sim P_\perp^2/\hat s$.

Below, we shall compute the dijet cross-section in the CGC effective theory, to leading order (LO) in $\alpha_s$ and to leading powers (LP) in the small ratios $K_\perp/P_\perp$ and $Q_s/P_\perp$. We shall thus find {\it TMD factorisation}: to the accuracy of interest, the cross-section is the product between  a {\it hard factor} (the same as for the corresponding processes at moderate $x$, see e.g. \cite{Qiu:2007ey}) and a {\it sea quark TMD}, for which the CGC approach offers explicit results from first principles. 

As we shall see, there are multiple such sea quark TMDs, depending upon the process. Yet, they share a universal structure determined by the colour flow of the hard partonic process. At the operator level, this universal structure reflects the various possible gauge links connecting bilocal quark field operators~\cite{Bomhof:2006dp}. Thus, even though there are several sea-quark TMDs, it is legitimate to speak of universality insofar as a given TMD enters several distinct processes sharing the same colour flow (in a manner analogous to gluon TMDs).
In the multicolour limit $N_c\to \infty$, all the emerging TMDs are built with only two ingredients: the sea quark TMD from the CGC calculation of SIDIS~\cite{Marquet:2009ca} and a colour dipole operator representing a Wilson loop which resums collinear gluon exchanges between the target and the initial-state and final-state partons in the hard process.

\section{Target versus projectile picture}

TMD (or collinear) factorisation is traditionally formulated in the target parton picture, where the small-$x$ quark that is knocked out by the collision is a constituent of the target --- the quark component of a sea quark-antiquark  ($q\bar q$) pair.  The CGC picture is however different. The target is described as a classical colour field (a ``shockwave'') representing  the small-$x$ gluons, but there are no explicit quark degrees of freedom: valence quarks are implicitly included as sources for the colour fields, but the sea quarks are absent\footnote{Fermionic background fields can be introduced via sub-eikonal corrections to the scattering.
This has been used to study quark TMD factorisation in dilute-dense collisions~\cite{Altinoluk:2023qfr,Altinoluk:2024dba,Altinoluk:2024tyx}. In this approach, the quark TMDs are not explicitly computed, but merely related to formal correlation functions of the fermionic background field.}.
In the CGC picture, the would-be {\it scattering} between a parton from the projectile and a sea  quark from the target is instead described as the {\it emission} of a soft {\it antiquark} by the projectile parton. This antiquark participates in the scattering with the target colour field, but it is not measured in the final state. Thus, the CGC calculation of dijet production requires the study of a three-parton Fock component of the light-cone (LC) wavefunction (WF) of the projectile (see the right graph in Fig.~\ref{fig:seaQ} for an example). To that aim,  the calculation should be performed in a Lorentz frame where the projectile is ultrarelativistic and in the projectile LC gauge. In the high-energy limit, the scattering can be computed in the eikonal approximation, meaning that projectile partons crossing the shockwave are not deflected, but merely suffer a colour precession described by a Wilson line in the appropriate representation of the colour group SU$(N_c)$.

Let us consider the DIS process in Fig.\,\ref{fig:seaQ} for definiteness. The same rationale will apply to the other dijet processes discussed in this Letter after the corresponding change of notation for the probe and final state particles. The CGC factorisation for this process, a.k.a. the colour dipole picture (CDP)~\cite{Kopeliovich:1981pz,Bertsch:1981py,Mueller:1989st,Nikolaev:1990ja}, is formulated in a frame where the virtual photon is a right-mover with 4-momentum $q^\mu=(q^+, q^- \!=\!-Q^2/2q^+, \bm{0}_\perp)$ with $q^+\gg Q$, while the nuclear target is a left mover with  $P_N^\mu\simeq \delta^{\mu-}P_N^-$ per nucleon. The photon LCWF is constructed in perturbation theory, in the LC gauge $A^+_a=0$. To compute quark-gluon ($qg$) dijet production to LO, one needs the 3-parton Fock state $q\bar q g$. Introducing LC longitudinal momenta $k_i^+\equiv z_i q^+$ and transverse momenta $\bm{k}_{i}$ for the three partons ($i=q,\bar q, g$), the dijet transverse relative momentum and imbalance are defined as $\Pt=z_g\bm{k}_{q}-z_q\bm{k}_{g}$ and $\Kt=\bm{k}_{q}+\bm{k}_{g}$, respectively. To simplify notations, we shall use $z$ and $\kt$ (instead of $z_{\bar q}, \bm{k}_{\bar q}$) for the unmeasured antiquark. For back-to-back dijets, one has $K_\perp\ll P_\perp$, while the longitudinal fractions $z_q$ and $z_g$ take generic values, which are not parametrically small. Throughout, we shall assume that $\PT$ is the hardest scale in the problem. The photon virtuality $Q^2$ can be comparable to $\PT^2$ or (much) lower than it. 

To uncover TMD factorisation for $qg$ dijets, we isolate the LP in the ratio $K_\perp/P_\perp$ 
and integrate out the kinematics $(z, \bm{k})$ of the unmeasured antiquark. As we shall see, this integration is controlled by  transverse momenta ${k}_{\perp}\sim K_\perp$ and small values $z\sim K_\perp^2/P_\perp^2\ll 1$ for the longitudinal momentum fraction. This can be understood from a formation time argument: in a hard process, the antiquark formation time $\Delta x^+\sim z q^+/{k}_{\perp}^2$ is comparable to that of the hard $qg$ pair:
\begin{align}\label{formtime}
\frac{z q^+}{K_{\perp}^2}\, \sim \,
\frac{z_q z_g q^+}{P_\perp^2}\ \Rightarrow\ z \sim \frac{K_\perp^2}{P_\perp^2}\,.
\end{align}
Thus, importantly, the unmeasured antiquark in the CGC picture (the same as the struck sea quark in the target picture) is {\it soft} both w.r.t.~to the projectile, $z\ll 1$, and w.r.t.~the target, $x\ll 1$. 
Being soft, it can indeed be ``transferred'' between the projectile and the target, as explained in~\cite{Hauksson:2024bvv,Iancu:2022lcw,Caucal:2024bae}.

This special kinematics for the three parton system also implies a special geometry for the collision in the transverse plane, as suggested by our drawing of the CGC  graph  in Fig.\,\ref{fig:seaQ}:  the hard $qg$ pair has a small size $r\sim 1/P_\perp\ll 1/Q_s$ and is separated from the soft antiquark
by  a  large distance $R\sim 1/K_\perp\gg r$. Accordingly, the incoming photon, the intermediate quark, and the hard $qg$ dijet are all aligned with each other: the respective transverse coordinates are roughly the same.  

In the back-to-back limit, the amplitudes of each process present an interesting correspondence between the CGC and the target picture. The CGC diagrams can be organized in terms of their ``bare" topologies which share the same splitting pattern but different positions for the shockwave insertion. Each CGC topology then corresponds to a diagram in the target picture. This correspondence, however, holds only to leading power in $1/\PT$  and within the same gauge choice. Then, the sum of all CGC diagrams sharing a given topology takes a factorised structure in which one of the factors encodes the dependence upon the kinematical variables $z_q,z_g,\Pt$ of the hard partonic subprocess, while the other one depends upon the variables $x,\Kt$ characterising the sea quark distribution in the target. This factorised structure at the amplitude level implies TMD factorisation for the di-jet cross-section, with  the same hard factor as in the target picture,  but with a more general expression for the sea quark TMD distribution, which also includes the effects of gluon saturation and of the high-energy evolution with decreasing $x$. The purpose of the three following sections is to unravel this correspondence between the CGC and target pictures in several dijet processes, and to present the result for the corresponding sea-quark TMDs.

\section{Quark-gluon dijet production in $eA$ DIS}

\begin{figure}
    \centering
    \includegraphics[width=1\linewidth]{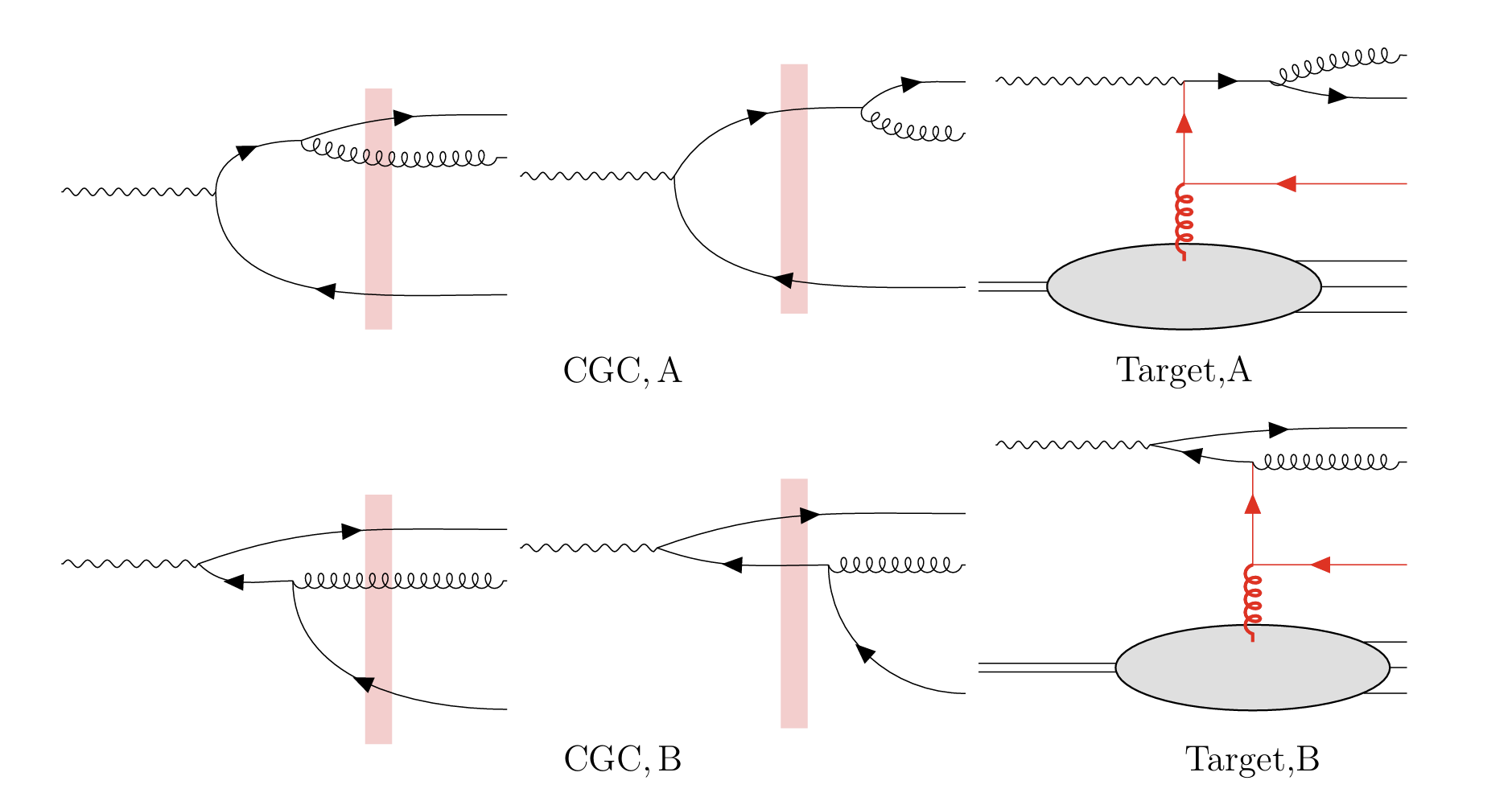}
    \caption{The correspondence between the CGC and the target picture diagrams for $qg$ production in DIS. On each row, the first two diagrams provide the set of diagrams with same topology. The third column shows the associated diagrams in the target picture. For more clarity, we show in black the part of a target-picture graph which encodes the hard partonic subprocess, and in red the part representing the formation of the sea quark via a gluon decay. }
    \label{fig:CGC_target_comp_DIS}
\end{figure}

Let us start by computing the cross-section for the CGC process
 $\gamma^*+A\!\to\! qg(\bar q)+X$. In general, there are two possible topologies: gluon emission by the quark, which we call topology A, or by the antiquark, which we refer to as topology B. The CGC diagrams associated with these two topologies are shown in Fig.~\ref{fig:CGC_target_comp_DIS} in the first two columns of each row, respectively. For a photon with transverse polarisation ($\gamma_T^*$), both topologies contribute on the same footing, but for a longitudinal one ($\gamma_L^*$), the first 
possibility turns out to be power-suppressed in the back-to-back limit. We present here the results for $\gamma_T^*$ and refer to~\ref{appA} for $\gamma_L^*$.

To LP in $K_\perp/P_\perp$, the sum of the amplitudes for the topology $A$ is obtained in the CGC as
 \begin{align}
    \Mcal_{qg(\bar q),A}^{\lambda\bar\lambda\sigma\bar\sigma,a}&=8ee_fg\sqrt{z_qz}q^+\delta^{\sigma,-\bar\sigma}\delta^{\sigma,-\lambda}\left[\delta^{\lambda\bar\lambda}+z_q\delta^{\lambda,-\bar\lambda}\right]\frac{\Pt\cdot\et^{\bar\lambda*}}{P_\perp^2}\nonumber\\
    &\times \int\frac{\der^2\lt}{(2\pi)^2}\frac{\ellt\cdot\et^{\lambda}}{\lt^2+z\left(Q^2+M_{qg}^2\right)}\mathcal{C}_1^a(\Kt+\lt,\kt-\lt)\,,\label{eq:Mqg-A-DIS}
\end{align}
where $M_{qg}^2\equiv (k_q+k_g)^2\simeq P_\perp^2/(z_q z_g)$, $\lambda=\pm1$ is the polarisation index of the virtual photon, $\bar\lambda$ and $a$ refer to the transverse polarisation and the colour of the produced gluon, $\sigma$ and $\bar\sigma$  are helicity indices for the quark and antiquark. Further, $g$ and $e$ denote the strong and electromagnetic charges, and $e_f$ is the fractional electric charge of a quark with flavour $f$. 
\eqn{eq:Mqg-A-DIS} involves the colour operator
\begin{align}
    \mathcal{C}_1^a(\qtone,\qttwo)=\int \der^2\xt\der^2\yt \ e^{-i\qtone\cdot \xt-i\qttwo\cdot\yt}\
    t^a[V_{\xt}V^\dagger_{\yt}-1]\,,\label{eq:C1-color}
\end{align}
with $V_{\xt}$ a light-like Wilson line in the fundamental representation of SU$(N_c)$ at fixed transverse coordinate $\xt$. The colour matrix $t^a$ refers to the gluon emission. By removing this matrix and replacing $g\to -ee_f$ in Eq.\,\eqref{eq:Mqg-A-DIS}, one gets the amplitude for $\gamma q\bar q$ production in DIS; hence  this paragraph also encompasses $\gamma$-jet correlations in DIS~\cite{Kolbe:2020tlq}. 

The Wilson lines in \eqn{eq:C1-color} describe the scattering of a $q\bar q$ colour dipole with a quark at $\xt$ and an antiquark at $\yt$. Since $\kt-\ellt$ is Fourier conjugate to $\yt$, one can physically interpret $\kt-\ellt$ as the transverse momentum transferred by the target to the antiquark, such that $\ellt$ is the transverse momentum of the antiquark before scattering off the shockwave. In the back-to-back limit, both CGC amplitudes in the first row of Fig.~\ref{fig:CGC_target_comp_DIS} share the same scattering operator, despite the different time orderings. Indeed, due to its small transverse size $r\sim 1/P_\perp\ll 1/Q_s$, the hard $qg$ pair effectively scatters in the same way as its parent quark at $\xt$. By comparison, the effective $q\bar q$ dipole which enters the amplitude \eqref{eq:Mqg-A-DIS} has a much larger size $R\equiv |\xt-\yt|\sim 1/\KT$ (since $k_\perp\sim \ell_\perp\sim \KT$; see below).

Consider now the topology B for the CGC diagrams, cf. the second row in Fig.~\ref{fig:CGC_target_comp_DIS}. The LP contribution entirely comes from the leftmost graph and involves again the colour matrix for the scattering of an {\it effective} $q\bar q$ dipole, with size $R\sim 1/\KT$:
 \begin{align}
    \Mcal_{qg(\bar q),B}^{\lambda\bar\lambda\sigma\bar\sigma,a}&=8ee_fg\sqrt{z_qz}q^+\delta^{\sigma,-\bar\sigma}\delta^{\sigma,\bar\lambda}\left[z_g\delta^{\lambda\bar\lambda}-z_q\delta^{\lambda,-\bar\lambda}\right]\frac{\Pt\cdot\et^{\lambda}}{P_\perp^2 + z_qz_gQ^2}\nonumber\\
    &\times \int\frac{\der^2\lt}{(2\pi)^2}\frac{\ellt\cdot\et^{\bar\lambda*}}{\lt^2+z\left(Q^2+M_{qg}^2\right)}\mathcal{C}_1^a(\Kt+\lt,\kt-\lt)\,.\label{eq:Mqg-B-DIS}
\end{align}
The other CGC graph is power-suppressed as it involves the scattering of the small $q\bar q$ dipole with size $r\sim 1/\PT$, which is suppressed by colour transparency.

To get the quark-gluon dijet cross-section, one must  square the sum of the amplitudes \eqref{eq:Mqg-A-DIS} and \eqref{eq:Mqg-B-DIS} and integrate the result over the phase-space of the unmeasured antiquark , i.e.~over $\kt$ and $z$. The mathematical structure of Eqs.\,\eqref{eq:Mqg-A-DIS}-\eqref{eq:Mqg-B-DIS} is such that the LP contribution to this phase space integral comes from the regime $k_\perp\sim \ell_\perp\sim \KT$ and $z\sim K_\perp^2/M_{qg}^2\sim \KT^2/\PT^2$, in agreement with the qualitative discussion around \eqn{formtime}. Taken separately, the square of the amplitude \eqref{eq:Mqg-A-DIS} for topology A gives a contribution to the hard factor identical to that obtained from the $s$-channel target-picture graph shown in the third column of the first row. Similarly, the square of the $B$-topology amplitude \eqref{eq:Mqg-B-DIS} provides the contribution to the hard factor corresponding to the square of the $t$-channel target-picture graph shown on the right of the second row. A similar discussion applies to the interference contributions to the hard factors in the two pictures. After combining direct and interference contributions, one finds the following result for the $qg$ dijet cross-section,
\begin{align}
    &\frac{\der \sigma^{\gamma_T^*A\to qg+X}}{\der^2\Pt\der^2\Kt\der z_q \der z_g}= \alpha_{s} \alpha_{em} e_f^2C_F \delta(1-z_q-z_g)\nonumber\\
    &\times\frac{ 2   z_q \left\{ \left[\Pt^2 + z_qz_gQ^2\right]^2 +z_g^2 (\Pt)^4 + z_q^4z_g^2Q^2\right\}}{\Pt^2 \left[z_qz_g Q^2 + \Pt^2 \right]^3}   xq^{(1)}(x,\Kt)\,,\label{eq:qg-final}
\end{align}
which exhibits TMD factorisation: it is the product of a hard factor  explicitly shown in \eqn{eq:qg-final} and the sea quark TMD $xq^{(1)}(x,\Kt)$ that will be exhibited shortly. The hard factor behaves like $1/P_\perp^4$ when $P_\perp^2\gtrsim Q^2$, showing that the overall result counts to leading-twist order, as expected.

The CGC result for the sea quark TMD relevant to this dijet process can be conveniently written as $xq^{(1)}(x,\Kt)\equiv\langle\mathcal{Q}(\Kt)\rangle_x $, where the brackets  $\langle ...\rangle_x$ refer to the CGC average over the target  colour fields   and  the sea quark operator $\mathcal{Q}$
\begin{align}
   \mathcal{Q}(\Kt)&\equiv\frac{N_c}{4\pi^4}\int\der^2\bt\der^2\qt \ \mathcal{D}_F^{(1)}(\qt,\bt) \nonumber\\
   &\times\left[1-\frac{\Kt\cdot(\Kt-\qt)}{K_\perp^2-(\Kt-\qt)^2}\ln\left(\frac{K_\perp^2}{(\Kt-\qt)^2}\right)\right]\label{eq:quark-TMD}\,,
\end{align}
involves the dipole operator~$\mathcal{D}_F^{(n=1)}(\qt,\bt)$, defined as (for arbitrary integer $n$, for later convenience)
\begin{align}
    \mathcal{D}_F^{(n)}(\qt,\bt)\equiv\int\frac{\der^2\rt}{(2\pi)^2}e^{-i\qt\cdot\rt}\left(\frac{1}{N_c} \Tr\left[V_{\bt+\rt/2}V^\dagger_{\bt-\rt/2}\right]\right)^n\,.
    \label{eq:dipole-def}
\end{align}
This operator depends upon the impact parameter $\bt$ of the $q\bar q$ dipole and the transverse momentum $\qt$ transferred by the target via the collision. Transverse momentum conservation implies that $\qt\!-\!\Kt$ is the momentum of the unmeasured antiquark. 

The $x$ dependence of the quark TMD $xq^{(1)}(x,\Kt)$ is controlled by the BK/JIMWLK~\cite{Balitsky:1995ub,Kovchegov:1999yj,Jalilian-Marian:1997qno,Jalilian-Marian:1997jhx,Kovner:2000pt,Weigert:2000gi,Iancu:2000hn,Iancu:2001ad,Ferreiro:2001qy} evolution of $\mathcal{D}_F^{(1)}$ down to $x=x_{qg}$, with $x_{qg}$ the target longitudinal fraction taken by the measured dijet. This is obtained from the condition of  $k^-$ conservation in the $\gamma^*q\to qg$ hard process as $x_{qg}= \left(M_{qg}^2+Q^2\right)/(2q^+P_N^-)$.

\begin{figure}[h]
   \centering
    \includegraphics[width=0.8\linewidth]{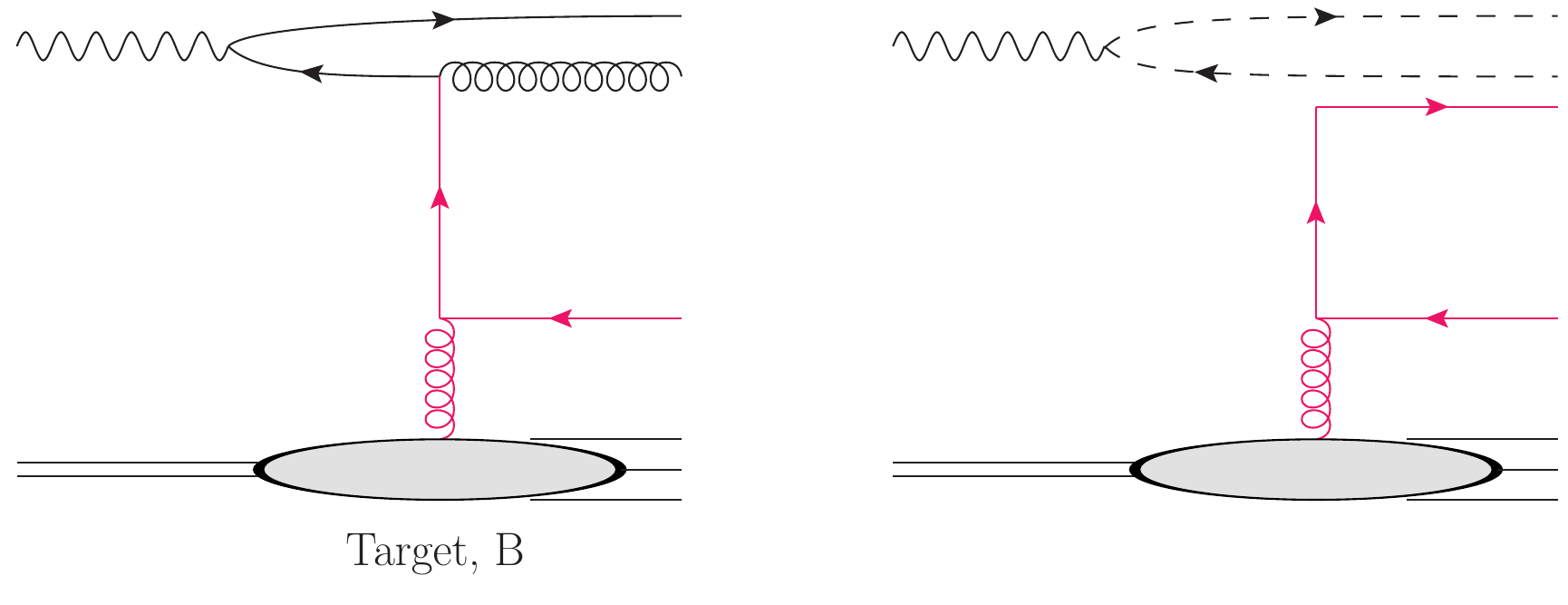}
    \caption{Left: target picture graph (topology B) for back-to-back quark-gluon dijets in $\gamma^*A$ collisions. Right: the associated colour flow at large $N_c$. }
    \label{fig:DISflow}
\end{figure}

The quark TMD $xq^{(1)}(x,\Kt)$ is identical to that obtained in the CGC calculation of SIDIS at small $x$~\cite{Marquet:2009ca,Xiao:2017yya}.
This is consistent with the fact that the colour flow is identical for both processes, due to the small transverse size of the hard dijet. This flow is illustrated in red in Fig.\,\ref{fig:DISflow} in the large $N_c$ limit for the topology B (the colour flow is identical in topology A). The dashed $q\bar q$ dipole which is disconnected from the quark TMD has a negligible transverse size $r\sim 1/P_\perp$, so it does not contribute to the colour flow.
\eqn{eq:quark-TMD} has a natural interpretation in the target picture \cite{Altinoluk:2024vgg}:
$\qt$ is the  transverse momentum of the parent gluon which generates the $q\bar q$ sea pair, while 
\beq\label{xG2}
 xG^{(2)}(x, \qt)\equiv \frac{q_\perp^2 N_c}{2\pi^2 \alpha_s} \int\der^2\bt \ \big\langle \mathcal{D}_F^{(1)}(\qt,\bt) \big\rangle_x,\eeq
represents the dipole unintegrated gluon distribution (UGD). As discussed in \cite{Dominguez:2010xd,Dominguez:2011wm}, this distribution plays the role of a  TMD for  small-$x$ processes which involve both initial-state and final-state interactions, i.e. which feature coloured partons on both the incoming and the outgoing legs. Of course, this is not the case in DIS, where all the coloured partons are outgoing. Yet, the dipole UGD is relevant in this context since the gluon distribution of the target is ``probed" by a  $q\bar q$ sea pair and the antiquark in this pair mimics an incoming quark in the initial state.
This suggests that the second line of \eqn{eq:quark-TMD}  times $\alpha_s/( 2\pi^2 q_\perp^2)$ can be interpreted as a {\it transverse momentum dependent  gluon-to-quark splitting function} (integrated over the splitting fraction $\xi$).  It indeed coincides with the off-shell splitting function employed in $k_T$-factorisation~\cite{Catani:1994sq,Ciafaloni:2005cg,Hautmann:2012sh,Hentschinski:2017ayz}. A more complete result, where the $\xi$-dependence is visible, can be found in~\cite{Xiao:2017yya,Altinoluk:2024vgg}.

\section{Photon-gluon jet in quark-initiated $pA$ collisions}

We now consider inclusive back-to-back photon-jet production in forward $pA$ collisions in the channel  $qA\to \gamma g(q)+X$. That is, the scattering is initiated by a quark from the proton (assumed to be on-shell prior to the collision) and the hadronic ``jet'' in the final state corresponds to a gluon --- so the final quark (with kinematical variables $z$ and $\kt$) is soft and unmeasured:
$k_\perp\sim K_\perp$ and $z\sim K_\perp^2/P_\perp^2\ll 1$. This process is therefore distinct from the production of a semi-hard photon and two hard jets studied in~\cite{Altinoluk:2018uax}. It should instead be included in the total NLO cross-section for isolated photon plus jet production in the correlation limit. The corresponding result for a photon plus a quark jet  involves the dipole UGD in \eqn{xG2}~\cite{Gelis:2002ki,Dominguez:2011wm}.

\begin{figure}[h]
    \centering
    \includegraphics[width=1\linewidth]{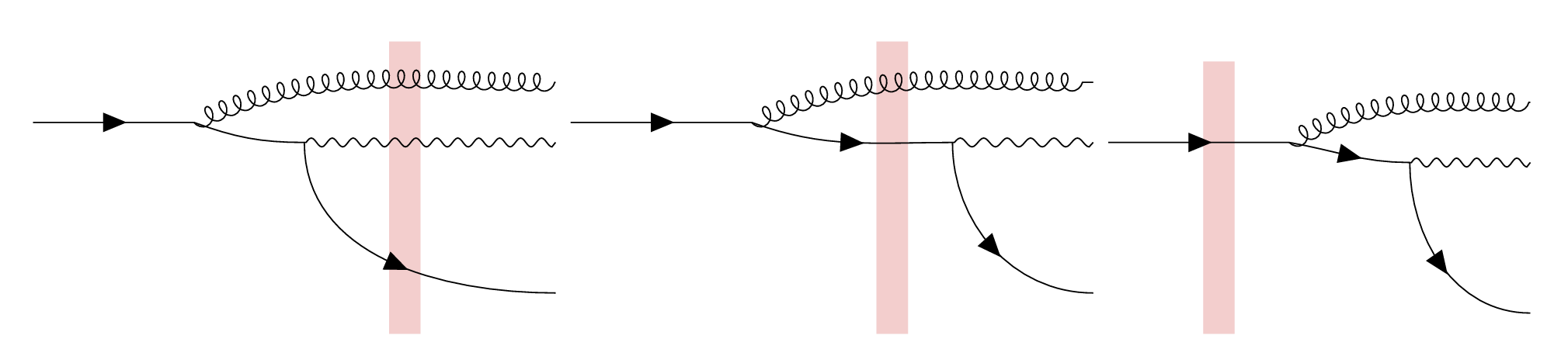}
    \caption{The three CGC diagrams contributing to the photon emission after gluon topology for $\gamma g$ production in pA collisions.}
\label{fig:CGC_vs_Target_gphoton_pA}
\end{figure}

\begin{figure}[h]
    \centering
   \includegraphics[width=0.8\linewidth]{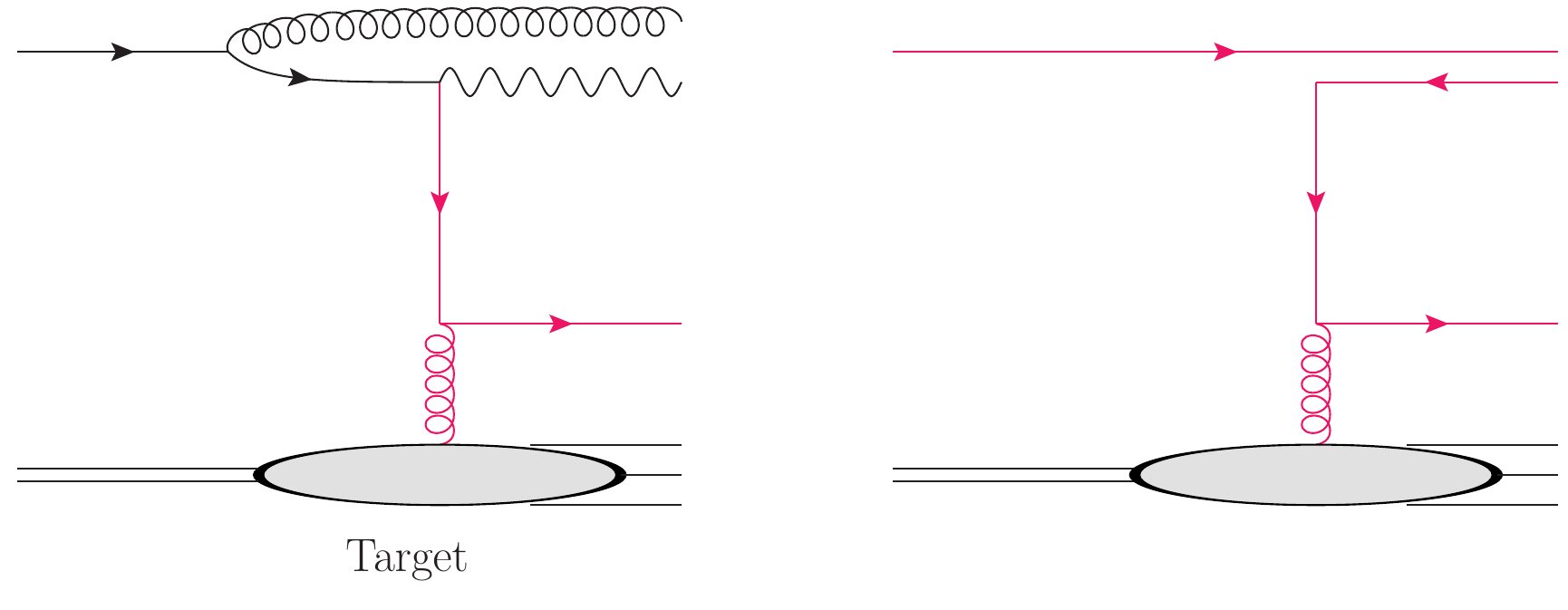}
    \caption{Left: one of the two target picture diagrams for $\gamma g$ jets in $pA$ collisions. Right: the associated
    colour flow at large $N_c$.     }
    \label{fig:pAflow}
\end{figure}

The CGC diagrams for $\gamma g$ production in $pA$ collisions involve two ``bare'' topologies: the photon is emitted either before, or after, the gluon. Each topology involves three graphs with different shockwave insertions.  In Fig.~\ref{fig:CGC_vs_Target_gphoton_pA}, we show the CGC diagrams where the photon is emitted after the gluon and the associated diagram in the target picture is shown in Fig.~\ref{fig:pAflow}-left. The corresponding diagrams for the second CGC topology are obtained by exchanging the gluon and the photon in all the graphs in Fig.~\ref{fig:CGC_vs_Target_gphoton_pA}.

As above, the total amplitude drastically simplifies after keeping only the LP in $ K_\perp/P_\perp$, and reads
\begin{align}
    \Mcal^{\bar\lambda\lambda\sigma\sigma'}_{\gamma g(q)}& =8ee_fg\sqrt{z}q^+\delta^{\sigma\sigma'}\delta^{\sigma,-\lambda}\left[\delta^{\lambda\bar\lambda}+z_\gamma\delta^{\lambda,-\bar\lambda}\right] \nonumber\\
    &\times\frac{\Pt\cdot\et^{\bar\lambda*}}{\Pt^2}\int\frac{\der^2\lt}{(2\pi)^2}\left[\frac{\kt\cdot\et^{\lambda*}}{\kt^2+zM^2_{\gamma g}}-\frac{\lt\cdot\et^{\lambda*}}{\lt^2+zM^2_{\gamma g}}\right]\nonumber\\
    &\times \mathcal{C}_2^a(\kt-\lt,\Kt+\lt)+(\gamma\leftrightarrow g)\,,\label{eq:gammajet-pA-amplitude}
\end{align}
where $\sigma,\sigma',\lambda,\bar \lambda$ respectively refer to the incoming and outgoing quark helicities and to the photon and gluon polarisation, $a$ to the colour of the gluon, and $M_{\gamma g}$ is the $\gamma$-jet invariant mass. The first (second) term inside the square brackets corresponds to gluon emission before (after) the shockwave. In the first case, the gluon participates in the scattering as well, so the respective amplitude involves the colour operator
\begin{align}
    \!\!\!\!\!\Ccal_2^a(\qtone,\qttwo)&=\int\der^2\xt\der^2\zt \ e^{-i\qtone\cdot\xt-i\qttwo\cdot\zt} V_{\xt}t^bU^{ab}_{\zt},
\end{align}
(with $U^{ab}_{\zt}$ an adjoint Wilson line) describing the scattering of quark-gluon system off the strong gluon field of the target. For the second term, the integral over $\lt$ yields a $\delta$-function enforcing $\xt=\zt$ which simplifies the colour operator as follows $V_{\xt} t^bU^{ab}_{\zt}\to t^aV_{\xt}$; as expected, this describes the scattering of the quark probe alone.

Squaring this amplitude and integrating over the kinematical variables $(z,\kt)$ of the unmeasured quark gives the inclusive $\gamma+$gluon-jet cross-section in the leading-twist approximation:
\begin{align}
    &\frac{\der \sigma^{qA\to \gamma g+X}}{\der^2\Pt\der^2\Kt\der z_\gamma \der z_g}=\delta(1-z_\gamma-z_g)\frac{\alpha_{\rm em}e_f^2\alpha_s}{N_c}\frac{(z_\gamma^2+z_g^2)}{\Pt^4}\nonumber\\
    &\hspace{1cm}\times\left[\frac{N_c}{2}xq^{(2)}(x,\Kt)-\frac{1}{2N_c} xq^{(1)}(x,\Kt)\right]\,,\label{eq:gamma-g-final}
\end{align}
with $x=x_{\gamma g}=M_{\gamma g}^2/(2q^+P_N^-)$. The first term, which is the only one to survive at large $N_c$,  features a new quark TMD operator at small $x$, defined as a convolution between our previous sea quark TMD operator $\mathcal{Q}(\Kt,\Deltat)$ and the dipole operator $\widetilde{\mathcal{D}}_F^{(1)}(\qt,\Deltat)$, where $\Deltat$ is the transverse momentum variable conjugated to the impact parameter $\bt$ via Fourier transform:
\begin{align}
xq^{(2)}(x,\Kt)&=\int\der^2\qt\der^2\Deltat \,
\left\langle
\widetilde{\mathcal{D}}^{(1)}_{F}\!\left(\qt, -\boldsymbol{\Delta}\right)
\mathcal{Q}\!\left(\Kt - \qt, \boldsymbol{\Delta}\right)
\right\rangle_{x}\,,\label{eq:qTMD2}
\end{align}
with
\begin{equation}
    \widetilde{\mathcal{D}}^{(n)}_{F}(\qt,\boldsymbol{\Delta})
=
\int \frac{\der^2 \bt}{(2\pi)^2}\,
     e^{-i\,\boldsymbol{\Delta}\cdot \bt}\,
     \mathcal{D}_F^{(n)}(\qt,\bt)\,,\label{eq:DqtDeltat}
\end{equation}
and
\begin{align}
    \mathcal{Q}(\Kt, \boldsymbol{\Delta})&= \frac{N_c}{2\pi^2}
\int\der^2\qt \,\widetilde{\mathcal{D}}^{(1)}_{F}(\qt, \Deltat)\nonumber\\
&\times\left\{
    2
    - \frac{\left(\Kt - \frac{\boldsymbol{\Delta}}{2}\right)
            \cdot \left(\Kt - \qt\right)}
           {\left(\Kt - \frac{\boldsymbol{\Delta}}{2}\right)^2
            - \left(\Kt + \qt\right)^2}
      \ln \!\left[
        \frac{\left(\Kt - \frac{\boldsymbol{\Delta}}{2}\right)^2}
             {\left(\Kt - \qt\right)^2}
      \right]\right.\nonumber\\
      &-\left. \frac{\left(\Kt + \frac{\boldsymbol{\Delta}}{2}\right)
            \cdot \left(\Kt - \qt\right)}
           {\left(\Kt + \frac{\boldsymbol{\Delta}}{2}\right)^2
            - \left(\Kt - \qt\right)^2}
      \ln \!\left[
        \frac{\left(\Kt + \frac{\boldsymbol{\Delta}}{2}\right)^2}
             {\left(\Kt - \qt\right)^2}
      \right]\right\}\,.
    \end{align}
Note that $\mathcal{Q}(\Kt,\boldsymbol{0})$ reduces to $\mathcal{Q}(\Kt)$ defined in Eq.\,\eqref{eq:quark-TMD}.

The target picture for this process and the associated colour flow at large $N_c$ are illustrated in Fig.\,\ref{fig:pAflow}. These figures make clear that the overall process involves  both initial and final state interactions. After replacing the gluon by a point-like $q\bar q$ pair (as appropriate at large $N_c$), the ensemble of these interactions can be resummed into a fundamental Wilson line extending from $x^+\to -\infty$ to $x^+\to \infty$, which factorises from the rest of the amplitude. This infinite Wilson line has a fixed transverse coordinate since the incoming quark and the outgoing gluon have roughly the same transverse positions.
Its product with the hermitian conjugate Wilson line from the complex conjugate amplitude yields the dipole operator visible in \eqn{eq:qTMD2}.  This can be recognised as a {\it Wilson loop} (since the transverse fields vanish at $x^+\to \pm\infty$), whose appearance was to be expected: such Wilson loop are well known to enter the gauge link structure of the operators defining quark TMDs in processes with both initial and final state interactions~\cite{Bomhof:2006dp}. 

\section{Gluon-quark dijets in gluon-initiated $pA$ collisions}

\begin{figure*}[t]
    \centering
    \includegraphics[width=0.62\textwidth]{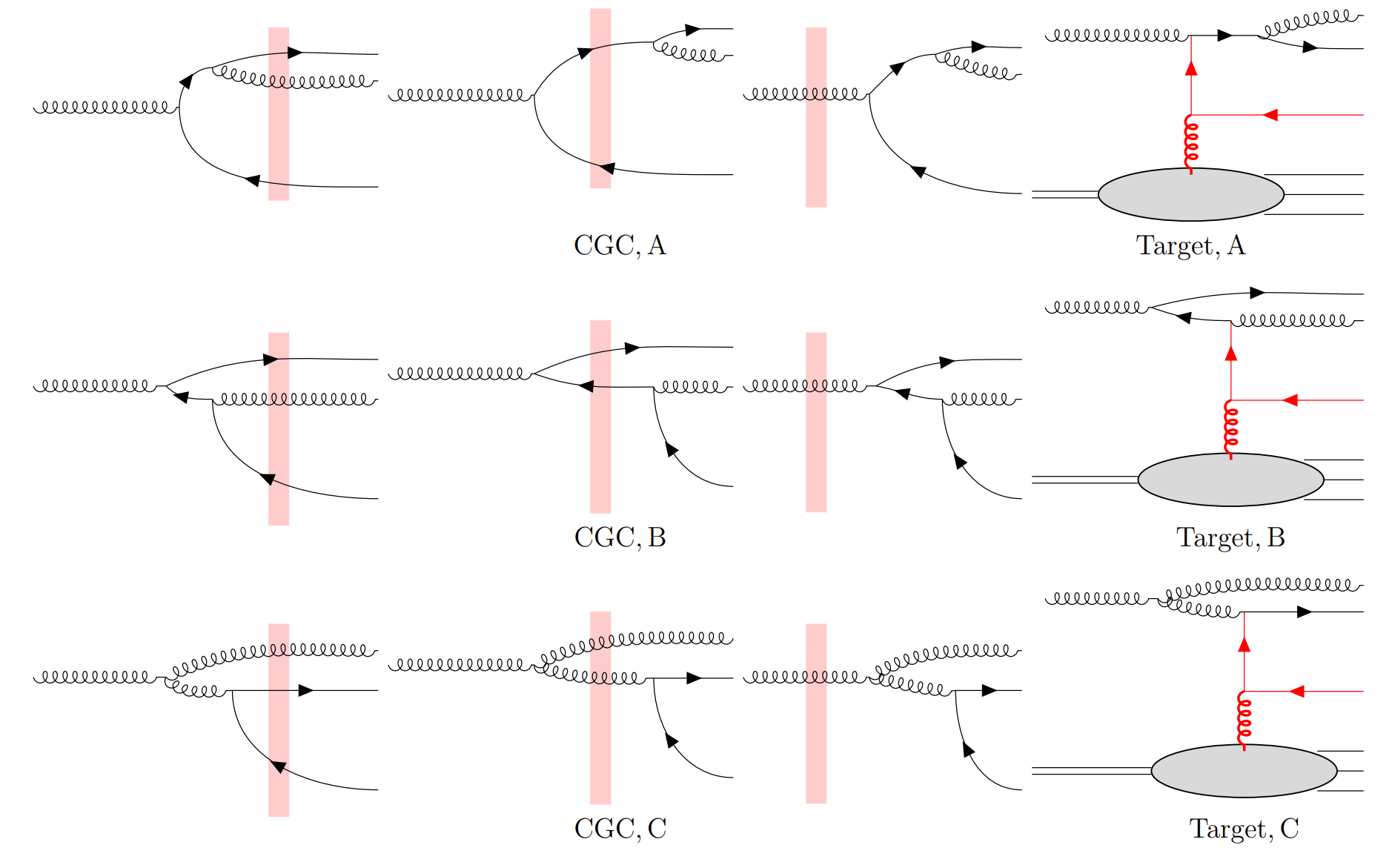}\hspace{1cm}
    \includegraphics[width=0.21\textwidth]{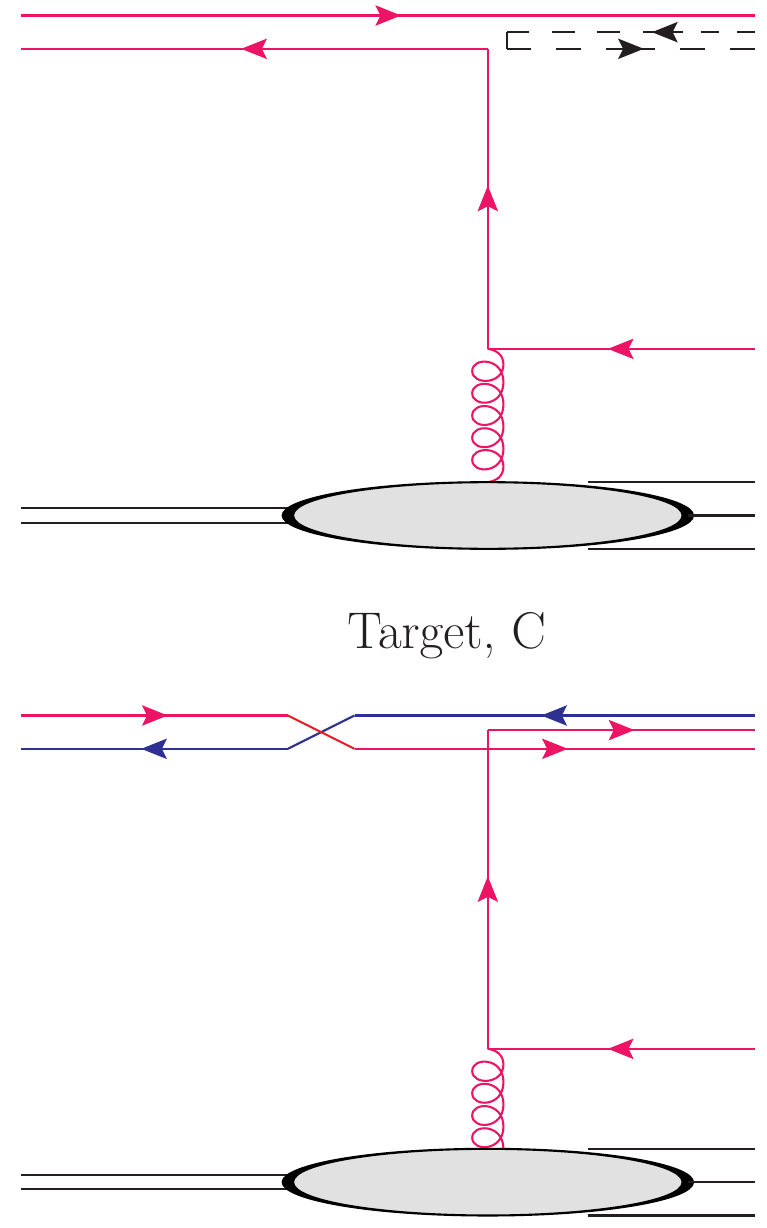}
    \caption{Left: the correspondence between CGC and target picture diagrams for $qg$ production in $pA$ collisions. The first three diagrams on each row correspond to the set of diagrams associated with the target diagram in the last column. Right: large-$N_c$ colour flow for the topology $C$ (gluon emission by the gluon). }
    \label{fig:CGC_vs_Target_qg_pA}
\end{figure*}

Our last example refers to gluon-quark dijets in $pA$ collisions
via the $g q\to gq$ hard process, which in the CGC calculation looks like $g+A\to qg(\bar q)+X$. The mapping between the CGC and the target picture elucidated in the two previous examples also works here. For $qg$ production in $pA$ collisions, the CGC diagrams can be grouped in terms of three topologies A, B and C: the gluon emission by a quark, an antiquark and a gluon, as illustrated in Fig.\,\ref{fig:CGC_vs_Target_qg_pA}. The first two topologies are similar to the DIS diagrams, but because the incoming parton has colour, we have an additional diagram with an interaction of the incoming gluon with the shockwave. In addition, we have a topology (C) where the final state gluon is emitted by the incoming gluon. The complexity of the colour structure of this process results in a non-trivial colour flow structure for each contribution. This colour flow can be inferred from the associated diagrams in the target picture, as shown in the third column of Fig.\,\ref{fig:CGC_vs_Target_qg_pA}.

We focus on the new topology C which involves the triple gluon vertex (last row in Fig.\,\ref{fig:CGC_vs_Target_qg_pA}) and which leads to the emergence of the new sea quark TMD $xq^{(3)}(x,\Kt)$, as we shall see. At large $N_c$, there are two patterns for the colour flow associated with this topology, shown in Fig.\,\ref{fig:CGC_vs_Target_qg_pA}-right: one where the quark component of the incoming gluon is transmitted to the outgoing gluon, and one where this component becomes the produced quark. In both cases, this  quark component gives rise to an infinite  Wilson line which generates a Wilson loop (or dipole correlator) in the cross-section.  In the second pattern  though, there is another such a Wilson line, associated with the antiquark component of the original gluon. Hence, the cross-section generated by this topology will feature two types of sea quark TMDs --- one  with a single Wilson loop, as shown in \eqn{eq:qTMD2},
 and another one involving two Wilson loops --- which in the transverse plane lie on top of each other.

 The colour flows depicted in Fig.\,\ref{fig:CGC_vs_Target_qg_pA}-right also reveals an interesting structure for the sea quark operator. In the first pattern, the (effective) colour flow corresponds to {\it initial-state interactions}; hence, the corresponding version of the operator $\mathcal{Q}(\Kt, \boldsymbol{\Delta})$ is the one that would naturally enter the Drell-Yan process ($q\bar q \to \gamma^*\to \ell^+\ell^-+X$) in $pA$~\cite{Xiao:2017yya}. For an unpolarised target, this coincides with the respective SIDIS operator in  \eqn{eq:quark-TMD}. In the second pattern, we recover the original SIDIS operator $\mathcal{Q}(\Kt, \boldsymbol{\Delta})$, with a colour flow encoding final-state interactions.

The total amplitude at LP in $K_\perp/P_\perp$ taking into account all topologies can be found in \ref{appB}. The final CGC result for the back-to-back $qg$ dijet cross-section reads (at large $N_c$)
\begin{align}
    &\frac{\der \sigma^{gA\to gq+X}}{\der^2\Pt\der^2\Kt\der z_q \der z_g}=\delta(1-z_q-z_g)\frac{\alpha_s^2(1+z_g^2)}{2z_q\Pt^4}\nonumber\\
    &\hspace{1.5cm}\times\left[xq^{(3)}(x,\Kt)+z_g^2 xq^{(2)}(x,\Kt)\right]\,,\label{eq:pA-qg-final}
\end{align}
where the quark TMDs are defined for generic integer $n\ge 1$ by the following generalisation of \eqn{eq:qTMD2}:
\begin{align}
xq^{(n)}(x,\Kt)&=\int\der^2\qt\der^2\Deltat \,
\left\langle
\widetilde{\mathcal{D}}^{(n-1)}_{F}\!\left(\qt, -\boldsymbol{\Delta}\right)
\mathcal{Q}\!\left(\Kt - \qt, \boldsymbol{\Delta}\right)
\right\rangle_{x}\,.\label{eq:qTMDn}
\end{align}
In the mean-field (or large $N_c$) approximation and for impact-parameter independent dipole amplitudes, it  simplifies to
\begin{align}
xq^{(n)}(x,\Kt)&=\int\der^2\qt
\left\langle
\mathcal{D}^{(n-1)}_{F}\!\left(\qt\right) \right\rangle_{x} xq^{(1)}(x,\Kt-\qt) \,,
\label{eq:qTMDn-meanfield}
\end{align}
where $ \mathcal{D}^{(n)}_{F}\!\left(\qt\right) $ is the impact parameter independent version of Eq.\,\ref{eq:dipole-def}. Interestingly, this expression matches the one conjectured in \cite{Xiao:2010sa} based on a scalar QED calculation.

\section{Numerical study}

We conclude this Letter with a  numerical study of the sea quark TMDs at small $x$ within the approximations employed in Eq.\,\eqref{eq:qTMDn-meanfield}, with particular emphasis on $xq^{(2)}$ and $xq^{(3)}$ that  have never been considered before. 
The basic building block is the dipole correlator,  for which we use the McLerran-Venugopalan (MV) model~\cite{McLerran:1993ka,McLerran:1993ni} $\langle \textrm{Tr}(V_{\br/2}V^\dagger_{-\br/2})\rangle=\exp[-\frac{1}{4}Q_s^2\br^2\ln((||\br||\Lambda)^{-1}+e)]$ with $\Lambda=0.24$ GeV.

The TMDs are then proportional to the overall transverse area of the target $S_\perp$. To study nuclear effects, we compare two types of targets: a dilute one (``proton'') with  saturation momentum  $Q_{s0}^2=0.2$ GeV$^2$ and a dense one (``nucleus'') with $Q_{s}^2=A^{1/3}Q_{s0}^2=1$ GeV$^2$, hence $A^{1/3}=5$.

\begin{figure}[h]
    \centering
    \includegraphics[width=0.45\textwidth,page=1]{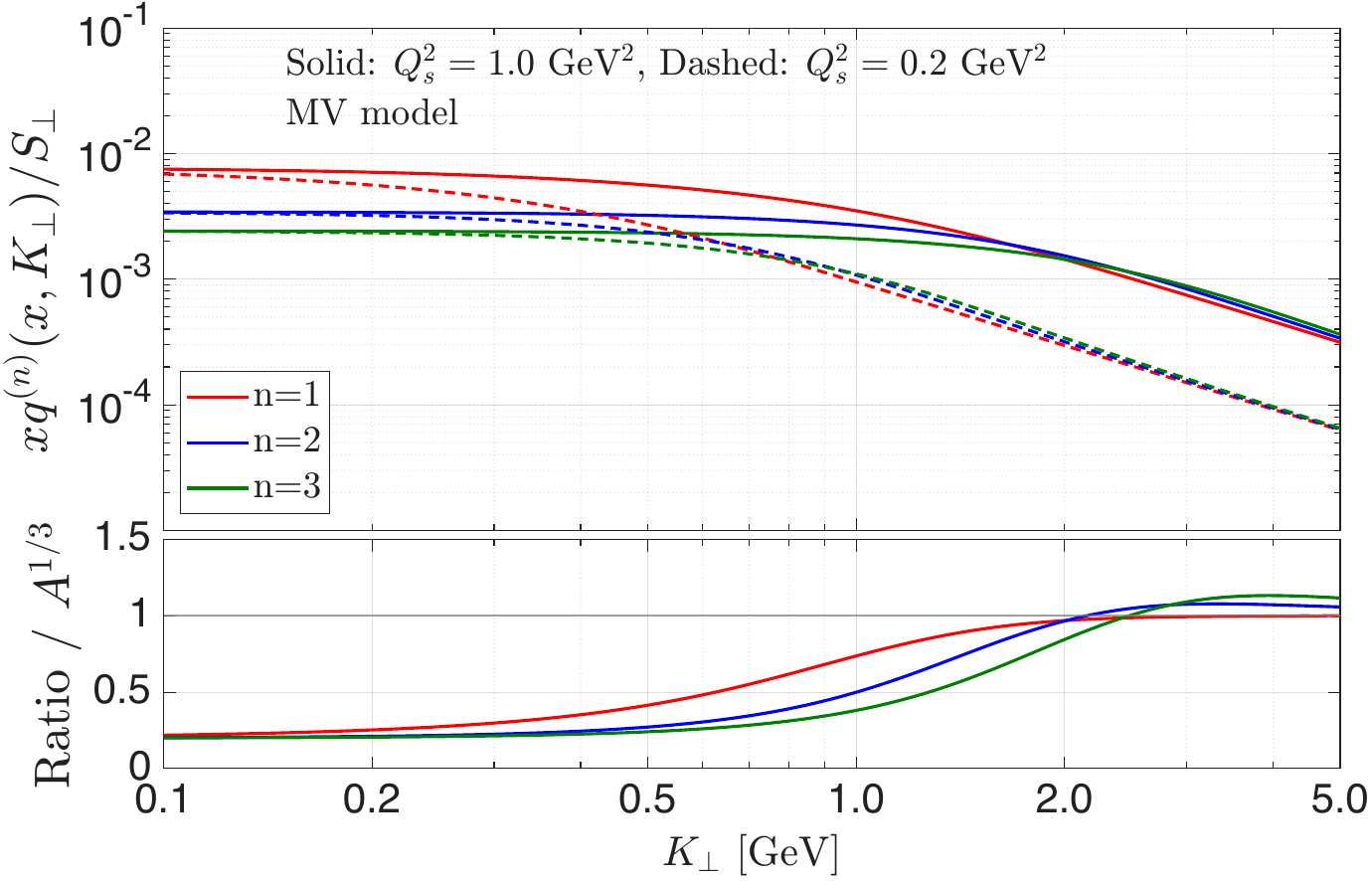}\hfill
    \caption{Small-$x$ quark TMD distributions (normalized by transverse area $S_\perp$ of the target) for a proton and a nucleus with $A^{1/3}=5$. Details in text.}
    \label{fig:qTMDs}
\end{figure}

Our results are presented  in Fig.\,\ref{fig:qTMDs}, with solid lines for a ``nucleus'' and dashed lines for a ``proton'', and for $n=1,2,3$. As visible in these plots, the TMDs display a universal behaviour at large $K_\perp$ featuring a perturbative power tail $1/K_\perp^2$. More precisely, in the MV model, the asymptotic behaviour of $xq^{(n)}(x,\Kt)/S_\perp$ at large $K_\perp$ is $\sim N_c/(24\pi^4)Q_s^2\ln(K_\perp^2/ \Lambda^2)/K_\perp^2$ and is therefore independent of $n$, as clear from Fig.\,\ref{fig:qTMDs}.
On the other hand, they saturate to different values at low momenta $K_\perp\lesssim Q_s$. The saturation region extends with increasing the power $n$. This is best seen in the nucleus over proton ratio TMDs (normalised by $A^{1/3}$) shown in the bottom plot: increasing $n$ effectively yields a larger  saturation scale and also a stronger deviation of the ratio from 1. This trend can be analytically understood from Eq.\,\eqref{eq:qTMDn}: for large $q_\perp\gg Q_s$, the dipole correlator is rapidly decaying, $ \langle \mathcal{D}_F^{(n)}(\qt)\rangle \propto 1/q_\perp^4$, while it is roughly flat in the saturation region $q_\perp^2\lesssim nQ_s^2$.
Hence the integral in Eq.\,\eqref{eq:qTMDn} is controlled by $q_\perp^2\lesssim nQ_s^2$ and this region extends 
when increasing $n$. 

Interestingly, the ratio for $n=2,3$ shows a Cronin-like enhancement which is not present for $n=1$. This can be explained as follows: unlike gluons \cite{Kharzeev:2003wz,Iancu:2004bx}, the total number of sea quarks per nucleon, as obtained by integrating the TMDs over $K_\perp$, is smaller in nuclei than in protons, meaning that the sea quark TMDs $xq^{(n)}(x,\Kt)$ suffer an overall suppression when comparing proton to the large nucleus case. This in turn prevents the usual Cronin peak due to broadening from appearing, unless the broadening is sufficiently large. This starts to be the case for $n\ge 2$, because of the aforementioned larger saturation window as $n$ increases. At sufficiently large $K_\perp$, the normalized nucleus-to-proton ratio eventually approaches one for any value of $n$, as can be analytically verified from the high-$K_\perp$ asymptotics of the sea-quark TMDs.

\section{Conclusion}

We have proven sea quark TMD factorisation for various  back-to-back dijet processes at high energy from the CGC. The sea quark TMDs have a universal structure, which is built from the fundamental sea quark TMD involved in SIDIS/Drell-Yan process at small $x$ and the quark-antiquark dipole correlator. This structure agrees with the gauge link structure generally expected for the quark TMDs, as controlled by the hard partonic subprocess and notably by the colour flow along the external legs not attached to the TMD \cite{Bomhof:2006dp}. These results pave the way towards the systematic incorporation of sea quark effects in the correlation limit of small-$x$ processes, a feature which has been largely overlooked in phenomenological works so far. For applications to the phenomenology, it is important to also take into account the 
additional, soft gluon radiation in the final state, which could strongly modify the back-to-back two-particle correlations. This can be done with the help of the Sudakov resummation method as developed in the CGC formalism~\cite{Mueller:2012uf,Mueller:2013wwa,Xiao:2017yya,Hatta:2021jcd,Taels:2022tza,Mukherjee:2023snp,Caucal:2023fsf,Caucal:2024bae,Caucal:2024nsb,Altinoluk:2024vgg,Caucal:2024vbv,Duan:2024nlr} or, alternatively, by coupling to parton showers~\cite{Zheng:2014vka,vanHameren:2016kkz,Cassar:2025vdp}.

\section*{Acknowledgements}
We are grateful to Xiaoxuan Chu, Larry McLerran, Yacine Mehtar-Tani, Al Mueller, Björn Schenke and Raju Venugopalan for discussions. Figures~\ref{fig:seaQ}, \ref{fig:DISflow}, \ref{fig:pAflow} and \ref{fig:CGC_vs_Target_qg_pA} have been created with JaxoDraw~\cite{Binosi:2003yf} We thank the France-Berkeley-Fund from University of California at Berkeley for support. PC is funded by the Agence Nationale de la Recherche under grant ANR-25-CE31-5230 (TMD-SAT). F.S. is supported by the Laboratory Directed Research and Development of Brookhaven National Laboratory and RIKEN-
BNL Research Center. Part of this work was conducted while F.S. was supported
by the Institute for Nuclear Theory’s U.S. DOE under
Grant No. DE-FG02-00ER41132.
This work is supported in part by the U.S. Department of Energy, Office of Science, Office of Nuclear Physics, under contract number DE-AC02-05CH11231 (F.Y.), under the umbrella of the Quark-Gluon Tomography (QGT) Topical Collaboration with Award DE-SC0023646. P.C., E.I. and F.S. are grateful for the support of the Saturated Glue (SURGE) Topical Theory Collaboration, funded by the U.S. Department of Energy, Office of Science, Office of Nuclear Physics.

\appendix

\section{Quark-gluon dijet in DIS: longitudinal photon}
\label{appA}
For a longitudinally polarised photon, the topology A in Fig.\,\ref{fig:CGC_target_comp_DIS} does not contribute at leading power in $K_\perp/P_\perp$, such that $\Mcal_{qg(\bar q),A}^{\bar\lambda\sigma\bar\sigma,a}=0$. The amplitude for the sum of CGC graphs with topology B yields
 \begin{align}
    \Mcal_{qg(\bar q),B}^{\bar\lambda\sigma\bar\sigma,a}&=-8ee_fg\sqrt{z_qz}q^+\delta^{\sigma,-\bar\sigma}\delta^{\bar\lambda\sigma}\frac{Q}{M^2_{qg}+Q^2}\nonumber\\
    &\times \int\frac{\der^2\lt}{(2\pi)^2}\frac{\lt\cdot \et^{\bar\lambda*}}{\lt^2+z\left(Q^2+M_{qg}^2\right)}\mathcal{C}_1^a(\Kt+\lt,\kt-\lt)\,\label{eq:Mqg-DIS}
\end{align}
in the back-to-back correlation limit. Squaring this amplitude, and integrating over the anti-quark phase-space $(z,\kt)$ gives the TMD factorised cross-section
\begin{align}
    \frac{\der \sigma^{\gamma_L^*A\to qg+X}}{\der^2\Pt\der^2\Kt\der z_q \der z_g}&=\delta(1-z_q-z_g)\frac{8\alpha_{\rm em}e_f^2\alpha_sC_FQ^2}{z_g\left(Q^2+M_{qg}^2\right)^3}  xq^{(1)}(x,\Kt)\,.\label{eq:qg-final-L}
\end{align}
for the $\gamma_L^*A\to qg+X$ process, where the hard factor is the same as in TMD factorisation at moderate $x$.

\section{Amplitudes for $g+A\to qg(\bar q)+X$}
\label{appB}
In this appendix, we provide the results for the leading power contribution to the amplitude and the differential cross-section for inclusive gluon-quark dijets in gluon-initiated $pA$ collisions. 

The total amplitude receives contributions from 9 CGC diagrams as shown in Fig.\,\ref{fig:CGC_vs_Target_qg_pA}, which can be classified into three types according to the gluon emitter: quark ($f=A$), antiquark $(f=B)$, or gluon $(f=C)$ parent. At LP in $K_\perp/P_\perp$ and $Q_s/P_\perp$ the total amplitude reads:
  \begin{align}
  \Mcal_{qg(\bar{q}),\sigma \bar{\sigma},i \bar{i}}^{\lambda\Bar{\lambda},ab}& = 8g^2 q^+ \sqrt{z_q z} \frac{ \Pt^k}{\Pt^2 } \sum_{f=A,B,C} \Gamma^{\lambda\bar{\lambda},jk}_{f,\sigma\bar{\sigma}} \nonumber\\
  &\times\int \frac{\der^2 \ellt}{(2\pi)^2} \Ccal^{ab}_{f,i \bar{i}}(\Kt+\ellt,\kt-\lt) \frac{\ellt^j}{z M_{qg}^2 + \ellt^2} \,,
\end{align}
where $\sigma,\bar{\sigma},\lambda,\bar \lambda$ respectively refer to the incoming and outgoing quark helicities and the incoming and outgoing gluon polarisations, and $i,\bar{i},a,b$ to the colors of quark, antiquark, and incoming and outgoing gluons. $j,k$ are transverse indices and $M_{qg}$ is the invariant mass of the $qg$-dijet. The corresponding splitting functions are:
\begin{align}
    \Gamma^{\lambda\bar{\lambda},jk}_{A,\sigma\bar{\sigma}}&=  \delta^{\sigma,-\bar\sigma} \delta^{\sigma,-\lambda} (\delta^{\lambda,\bar{\lambda}}  + z_q \delta^{\lambda,-\bar{\lambda}} ) \et^{\bar{\lambda}*,k}\et^{\lambda,j} \,, \\
    \Gamma^{\lambda\bar{\lambda},jk}_{B,\sigma\bar{\sigma}}&=   \delta^{\sigma,-\bar\sigma} \delta^{\sigma,\bar{\lambda}} (z_g \delta^{\lambda, \bar{\lambda}} - z_q \delta^{\lambda,-\bar{\lambda}}) \et^{\lambda,k} \et^{\bar{\lambda},j*} \,, \nonumber \\
    \Gamma^{\lambda\bar{\lambda},jk}_{C,\sigma\bar{\sigma}}&= z_g \delta^{\sigma,-\bar\sigma} \left[\delta^{\bar{\lambda},\sigma} \et^{\lambda,k} - \frac{\delta^{\lambda,\bar{\lambda}} \et^{\sigma,k}}{z_q} - \frac{\delta^{\lambda,-\sigma} \et^{\bar{\lambda},k*}}{z_g} \right] \et^{\sigma,j*} \,, \nonumber
\end{align}
and the color operators
\begin{align}
    \Ccal^{ab}_{A,i \bar{i}}(\qtone, \qttwo) &= \int\der^2\xt\der^2\yt \ e^{-i\qtone\cdot \xt} e^{-i\qttwo\cdot \yt}  \left[t^b V_{\xt}t^a V^\dagger_{\yt} - U^{ca}_{\xt} t^b t^c  \right] \,, \nonumber \\
    \Ccal^{ab}_{B,i \bar{i}}(\qtone,\qttwo) &= \int\der^2\xt\der^2\yt \  e^{-i\qtone\cdot \xt} e^{-i\qttwo\cdot \yt} \left[ U^{bc}_{\xt} V_{\xt} t^a t^c V^\dagger_{\yt} - U^{ca}_{\xt}  t^c t^b  \right] \,, \nonumber \\
   \Ccal^{ab}_{C,i \bar{i}}(\qtone, \qttwo) & = \Ccal^{ab}_{A,i \bar{i}}(\qtone, \qttwo) - \Ccal^{ab}_{B,i \bar{i}}(\qtone, \qttwo) \,.
\end{align}
The contributions coming from gluon emission from quark and gluon emission from antiquark are analogous to those in the $qg$ jets in DIS (in the limit $Q^2 \to 0$). The splitting functions $\Gamma_A$ and $\Gamma_{B}$ are the same, while the color operators are different due to the color of the incoming gluon. The contribution from gluon emission from gluon is a new feature of this channel. Note that the color operator for the gluon emission from gluon can be written in terms of those for gluon emission from quark and antiquark as anticipated from our heuristic discussion at large $N_c$ (cf.~Fig.\,\eqref{fig:CGC_vs_Target_qg_pA}-right).
\bibliographystyle{elsarticle-num}
\bibliography{refs}






\end{document}